# Intrinsic mechanism for spectral evolution in single-frequency Raman fiber amplifier

Wei Liu,[1] Pengfei Ma,[1,2,3] Yu Miao,[1] Pu Zhou,[1,2,3,*] Hanshuo Wu[1] and Zongfu Jiang[1,2,3]

[1]*College of Optoelectronic Science and Engineering, National University of Defense Technology, Changsha, Hunan, 410073, People's Republic of China*

[2]*Hunan Provincial Key Laboratory of High Energy Laser Technology, National University of Defense Technology, Changsha, Hunan, 410073, People's Republic of China*

[3]*Hunan Provincial Collaborative Innovation Center of High Power Fiber Laser, National University of Defense Technology, Changsha, Hunan, 410073, People's Republic of China*

*\*zhoupu203@163.com*

**Abstract:** In this work, the spectral evolution properties in single-frequency Raman fiber amplifier (RFA) with different pump manners are analyzed theoretically for the first time based on the gain dynamics. The analysis of gain dynamics reveals that the walk-off effect in counter-pumped manner produces a natural low-pass filter in single-frequency RFA. When applying rare-earth doped fiber lasers as the pump source, the strong temporal fluctuations in the pump source lead to spectral broadening in co-pumped manner, while the natural low-pass filter in countered-pumped case can still ensure single-frequency operation. Furthermore, applying temporal stable laser, such as single-frequency fiber laser or narrow band fiber laser spectral broadened by phase modulation technique, as the pump source would be superior for high-performance single-frequency RFA for the both two pump manners.

## 1. Introduction

Raman fiber laser (RFL) is a specific type of fiber laser based on the gain of stimulated Raman scattering (SRS) effect. Due to its arbitrary output wavelength by choosing the pump-laser wavelength, high quantum efficiency and outstanding cascaded amplification capability, it could provide almost any wavelength in near-infrared region and has great potential for high-power and wide wavelength range amplification [1, 2]. In its developments, single-frequency RFA is discovered to be a promising in nonlinear frequency conversion to obtain single-frequency fiber lasers at some special wavelength. For example, single-frequency RFL operating at 1120-1150 nm can be frequency doubled to generate laser emission at 560-575 nm, which is widely used in medical science, biological imaging and so forth [3-5]. Single-frequency RFL operating at 1178 nm can be frequency doubled to generate laser emission at 589 nm, which is widely used as guide star for adaptive optics [6-8].

Previous experimental study had been shown interesting and special spectral broadening properties in single-frequency RFA with different pump manners [9]. Specifically, single-frequency operation could be achieved in counter-pumped RFA while significant sideband spectral broadening is observed in a co-pumped manner. These spectral broadening phenomena in single-frequency RFA are quite different from that in common rare-earth doped single-frequency or multi-longitudinal mode fiber lasers [10-12]. In the following publications, high-power single-frequency RFAs with different operating wavelength are experimentally fulfilled and reported by directly using counter-pumped structure while little publication is focused on searching the physical origin of the spectral broadening differences in single-frequency RFA.

In single-frequency fiber laser, the spectral distribution is mainly characterized by the spectral density of frequency noise. The spectral broadening effect is commonly just attributed to the frequency fluctuations [13, 14]. However, it is worth noting that those results are based on ideal single-frequency laser without any intensity fluctuations. When there are intensity fluctuations in single-frequency laser, the corresponding spectral properties might be quite different. In this paper, we aim to explore the spectral evolution properties in single-frequency RFA by simultaneously considering the frequency and intensity fluctuations. The gain dynamics in a single-frequency RFA using the coupled amplitude equations is employed to express intensity noise transformation of pump-to-pump, pump-to-signal and signal-to signal, which accounts for both the pump-to-signal noise transformation and the nonlinear propagation of the signal light. Theoretical results show that the gain dynamics are rather different in co-pumped and counter-pumped configuration for single-frequency RFA. Different from co-pumped configuration, there exists a natural low-pass filter to weaken intensity noise transferred from the pump fluctuations and avoid sideband broadening in counter-pumped configuration. Furthermore, the pump-configurations would have little influence on the performance of single-frequency RFA when applying temporal stable laser as the pump source, such as single-frequency fiber laser or narrow band fiber laser spectral broadened by phase modulation technique.

## 2. Gain dynamics analysis in a single-frequency RFA

In the traditional analysis of gain dynamics, the nonlinear effects, such as self-phase modulation (SPM), cross-phase modulation (XPM) or group velocity dispersion (GVD), would not induce pump-to-signal power transformation. Meanwhile, the time-dependent rate equations or the coupled power equations are capable to analyze the Raman gain, pump depletion and saturation effect, thus only the power transformation process is considered [15, 16]. To understand the output intensity noise properties transferred from the pump light, it is useful to apply the pump modulation transfer function in the frequency domain. To obtain the pump modulation transfer function, we impose a small cosine modulation of the pump power:

$$P_p(t) = P_p^0 \left[1 + \delta_0 \cos(2\pi f t)\right] \quad (1)$$

Here, $P_p^0$ is the initial pump power, $\delta_0$ is the modulation depth and $f$ is the modulation frequency. Then, the output signal power and the corresponding magnitude of pump modulation transfer function $|T(f)|$ could be expressed as:

$$P_s(t) = P_s^0 * \left[1 + \delta_1(f) \cos(2\pi f t + \phi)\right] \quad (2)$$

$$|T(f)| = \delta_1(f)/\delta_0 \quad (3)$$

Here, $P_s^0$ is the steady state output signal power and $\phi$ is the phase delay.

To explore the spectral characteristics of single-frequency RFA, we need to consider both the properties of the intensity and phase noise, while not just the intensity noise. Similar as shown in traditional nonlinear optics [17], we consider the phase noise transfer property in single-frequency RFA by using the nonlinear Schrödinger equation:

$$\begin{cases} \pm\dfrac{\partial A_p^\pm}{\partial z} + \dfrac{1}{v_{gp}}\dfrac{\partial A_p^\pm}{\partial t} + \dfrac{i\beta_{2p}}{2}\dfrac{\partial^2 A_p^\pm}{\partial t^2} + \dfrac{\alpha_p}{2}A_p^\pm = i\gamma_p\left[\left|A_p^\pm\right|^2 + (2+\delta_R - f_R)\left|A_s\right|^2\right]A_p - \dfrac{g_p}{2}\left|A_s\right|^2 A_p^\pm \\ \dfrac{\partial A_s}{\partial z} + \dfrac{1}{v_{gs}}\dfrac{\partial A_s}{\partial t} + \dfrac{i\beta_{2s}}{2}\dfrac{\partial^2 A_s}{\partial t^2} + \dfrac{\alpha_s}{2}A_s = i\gamma_s\left[\left|A_s\right|^2 + (2+\delta_R - f_R)\left|A_p^\pm\right|^2\right]A_s - \dfrac{g_s}{2}\left|A_p^\pm\right|^2 A_s \end{cases} \quad (4)$$

Here, + and – denote for the co-pump and counter-pump cases, $A$ is the envelope of the optical field; index $p$ and $s$ stands for pump wave and Raman stokes wave, respectively. $v$ is the group velocity, $\beta_2$ is the second order dispersion coefficient, $\alpha$ is the loss coefficient, $\gamma$ is the nonlinear Kerr coefficient, $\delta_R$ is the Raman-induced index changes. $f_R$ represents the fractional contribution of the delayed Raman response to nonlinear polarization, and $g$ is the Raman gain coefficient.

Thus, the full dynamical process of single-frequency RFA could be described by the combination of the pump light property and the nonlinear propagation process. Then, both the frequency and intensity fluctuations in single-frequency RFA could be considered and analyzed simultaneously. Due to that our main purpose is to investigate the spectral properties while not power scaling limitations, so we assume that single-frequency RFA is operated below the stimulated Brillouin effect (SBS) threshold for simplicity. Of course, the SBS effect can be also considered by incorporating the interaction process of acoustic and optical fields.

Numerical simulation method is used to analyze the spectral broadening mechanisms in single-frequency RFA with different pump manners here. The major simulation parameters are shown as follows: $\lambda p = 1070nm$, $\lambda s = 1120nm$, $v = 2\times10^8 m/s$, $\alpha = 0.015 dB/m$, $\beta 2 = 20 ps^2/km$, $\gamma = 10 W^{-1}/km$, $fR = 0.245$, $gs = 4.2\ W^{-1}/km$, $gp = 4.4\ W^{-1}/km$ and $\delta 0 = 0.1$. The seed power is 0.1 W, the pump power is 10 W and the fiber length is 20 m. In order to describe the practical interactions process between the pump wave and the signal wave with different pump manners, we applied the parallelizable, bidirectional finite-difference time domain method in the simulation here [18].

Figure 1 illustrates the simulated magnitude of the pump modulation transfer function with different pump manners. As shown in Fig. 1, the pump modulation transfer function behaves as an all-pass filter in co-pumped manner. While in counter-pumped manner, the overall curve of the magnitude satisfies the properties of a low-pass filter and the cut-off frequency is about 3 MHz here (the blue line and diamonds in Fig. 1).

This result could be understood by the pump-signal walk-off effect in counter-pumped manner. In counter-pumped manner, the pump light and the signal light propagate on the opposite directions in the fiber, thus there exists time delay between signal light and pump light along the fiber. Accordingly, only the average intensity of the pump light over the propagation time along the fiber would impact the temporal properties of the signal light.

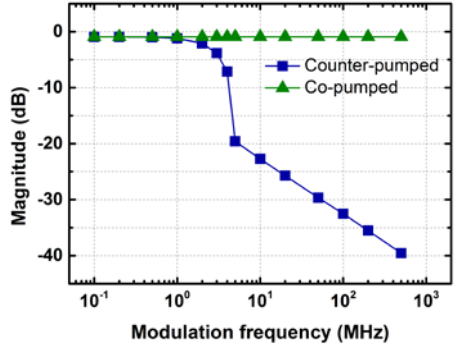

Fig. 1. The simulated magnitude of the pump modulation transfer function in RFA.

Based on the above analysis, the magnitude of the pump modulation transfer function in counter-pumped manner can be calculated analytically at certain simplifications. When neglecting the pump depletion and the fiber loss, the Raman amplification process could be simply expressed as:

$$\frac{dP_s}{dz} = g_s P_p P_s \tag{5}$$

Substituting Eq. (1) into Eq. (5), the output signal power and the magnitude of the pump modulation transfer function could be derived as:

$$\begin{aligned} P_s(t) &= P_{seed} \cdot \exp\left(\int_0^L g_s P_p(z) dz\right) \\ &= P_{seed} \cdot \exp\left(\int_0^T v g_s P_p(t+2t') dt'\right) \\ &= P_{seed} \cdot \exp\left(g_s P_p^0 L\right) \cdot \exp\left[v g_s \delta_0 P_p^0 \int_0^T \cos\left(2\pi f(t+2t')\right) dt'\right] \\ &\approx P_s^0 \cdot \left\{1 + v g_s \delta_0 P_p^0 / (4\pi f)\left[\sin(2\pi ft)(\cos(4\pi fT) - 1) + \cos(2\pi ft)\sin(4\pi fT)\right]\right\} \\ &= P_s^0 \cdot \left[1 + v g_s \delta_0 P_p^0 / (4\pi f) \cdot |2\sin(2\pi fT)| \cdot \cos(2\pi ft + \phi)\right] \end{aligned} \tag{6}$$

$$|T(f)| = g_s P_p^0 L |\sin(2\pi fT)|/(2\pi fT) \tag{7}$$

Here, $T$ is the propagation time of the signal light along the fiber. We may notice that the shape of the transfer function satisfies the absolute value of Sinc function and the period of the sine function part is equal to the propagation time of the signal light in the fiber. Figure 2 compares the simulated magnitude of the pump modulation transfer function with analytical results through Eq. (7). As shown in Fig. 2, the analytical results agree quite well with the simulation results.

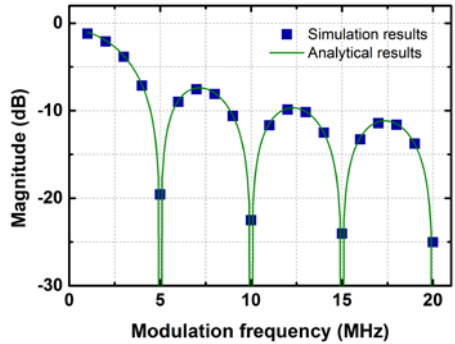

Fig. 2. Comparison between the simulated magnitude of the pump modulation transfer function with the analytical results.

In summary, the pump modulation transfer function in RFA behaves as an all-pass filter in co-pumped manner and the pump-signal walk-off effect in counter-pumped manner produces a natural low-pass filter. Thus, the pump fluctuations at high Fourier frequency would be filtered in counter-pumped manner during amplification. Besides, the shape of the low-pass filter is close to the absolute value of Sinc function and the period of the sine function part is equal to the propagation time of the signal light in the fiber.

## 3. Spectral broadening in a single-frequency RFA

As is illustrated in the above section, the intensity noise in the pump light would transform to the signal light in single-frequency RFA. Thus, the temporal properties of the pump light are quite important in the performance of the single-frequency RFA. For core-pumped RFA, the typical pump sources are rare-earth doped fiber lasers, which induces strong temporal fluctuations in picosecond scale [19]. The spectral and temporal characteristics of this source can be calculated through combined simulation of the rate equations and nonlinear Shrödinger equations [20-22]. Here, we apply a simpler way to character the temporal fluctuations through the polarized thermal radiation model [23]:

$$\tilde{A}_p(\omega) \propto \exp\left(-2In(2)\frac{\omega^2}{\Omega_L^2}\right)\exp(i\varphi(\omega)) \qquad (8)$$

Where $\Omega_L$ is the spectral width and the random spectral phase $\varphi(\omega)$ obeys the uniform probability distribution between $-\pi$ to $\pi$.

Substituting the Eq. (8) into Eq. (4), we could simulate the temporal and spectral characteristics of single-frequency RFA with the typical multi-longitudinal mode fiber oscillator as the pump source, and the same simulation parameters are used here as in the above section, except that the seed power is 2 W here.

Figures 3(a) and 3(b) illustrate the simulated temporal evolution and corresponding optical spectra in counter-pumped single-frequency RFA. As shown in Fig. 3(a), the signal light in counter-pumped manner would become relative stable during amplification, since the pump fluctuations at high Fourier frequency would be filtered in this case. Thus, the single-frequency operation can be preserved in counter-pumped manner (shown in Fig. 3(b)). Figures 3(c) and 3(d) illustrate the simulated temporal evolution and corresponding optical spectra in co-pumped single-frequency RFA. As shown in Fig. 3(c), the signal light in co- pumped manner would induce strong temporal fluctuations in the nanosecond scale (the average power is about 4 W here). This property is consistent with the temporal property of the pump light. Meanwhile, when the pump fluctuations are integrally transferred into the signal, only the background noise in the sideband is amplified (shown in Fig. 3(d)), which corresponds the phenomenon of the spectral broadening in measurement of the optical spectral analyzer.

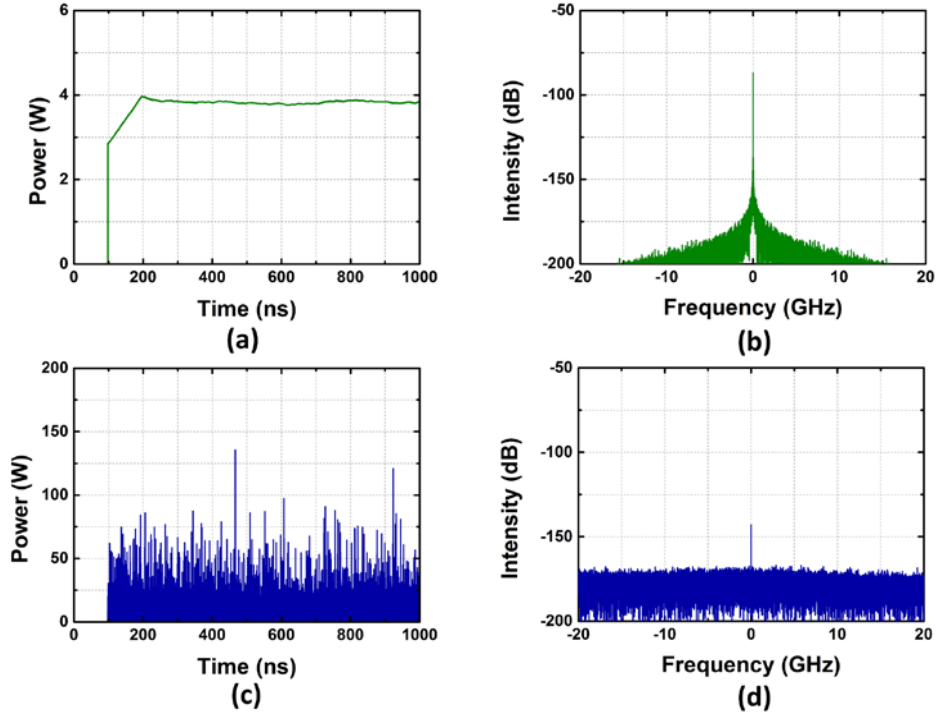

Fig. 3. The simulated temporal evolution and corresponding optical spectra for the single-frequency RFA: (a) (b) in counter-pumped manner; (c) (d) in co-pumped manner.

Although only the pump fluctuations at low Fourier frequency are transferred into the signal in counter-pumped manner, it would still lead to the amplification of the background noise in the sideband and degrade the signal-to-noise ratio in further amplification. A direct idea to avoid this phenomenon is to apply the temporal stable laser source, such as single-frequency fiber lasers or the phase-modulated single-frequency laser sources.

Figures 4(a) and 4(b) illustrate the simulated temporal evolution and corresponding optical spectra in single-frequency RFA with the phase-modulated single-frequency fiber laser. The modulation frequency is set to be 100 MHz with the modulation depth that can be produced three discrete single-frequency peaks, which ensures about three times SBS threshold enhancement compared to the single-frequency fiber laser [24]. The temporal characteristics of the signal light is quite stable (shown in Fig. 4(a)) and the single-frequency operation can be preserved (shown in Fig. 4(b)) in the both pump manners. Thus, applying the temporal stable laser source as the pump light would be superior for high-performance single-frequency RFA.

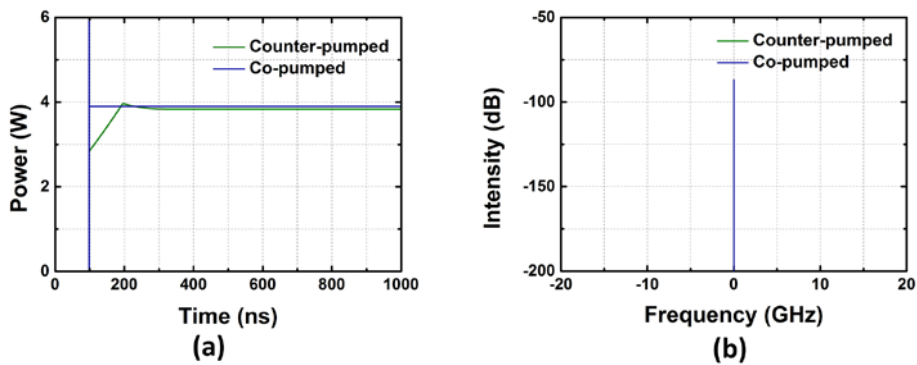

Fig. 4. The simulated temporal evolution and corresponding optical spectra with the phase-modulated single-frequency light: (a) the temporal evolution; (b) the corresponding optical spectra.

## 4. Conclusions

In this work, the intrinsic mechanism for spectral evolution in single-frequency RFA are analyzed theoretically based on the gain dynamics. When applying the rare-earth doped fiber lasers as the pump source, we show that the natural low-pass filter induced by walk-off effect in counter-pumped manner single-frequency RFA can protect the spectral distribution from broadening in sideband. However, strong temporal fluctuations are still existed and will lead to the spectral broadening in co-pumped manner. Further analysis shows that applying the temporal stable laser as the pump source is a robust way to avoid spectral broadening both in counter and co-pumped manners.